# Size distribution modelling of secondary atomization in spray of plain-jet airblast atomizer with finite stochastic breakup model

Maziar Shafaee[1*], Mohammad Hosein Sabour[2], Armin Abdehkakha[3], Abbas Elkaie[4]

[1]Proffessor Assistant, Faculty of New Sciences and Technologies, Tehran University, mshafaee@ut.ac.ir

[2]Proffessor Assistant, Faculty of New Sciences and Technologies, Tehran University, saburmh@ut.ac.ir

[3]Master student, Faculty of New Sciences and Technologies, Tehran University, arminabdeh@ut.ac.ir

[4]Master student, Faculty of New Sciences and Technologies, Tehran University, a.elkaie@ut.ac.ir

An experimental investigation of secondary atomization of spray in plain-jet airblast atomizers with circular array of six liquid jets, is studied. Cross flow of air with controlled velocity and pressure is applied to transverse jets. The liquid jets (Re=12000) are injected into air flow ($We_g$ of 1250-3000) analyzed experimentally for different injection diameters of 0.8-1.6 mm. Particle size distribution measurement is carried out by Malvern Mastersizer X in fully atomized region of the atomizer. The droplet size distribution in secondary or final atomization stage is also modelled by Finite stochastic Breakup model, which requires 4 parameters to be defined i.e. initial droplet diameter, maximum stable droplet size, minimum mass ratio and droplet breakup probability. The modeling approximation is in very good agreement with experimental result. This remarkable consistency between model and experiment is quite useful in terms of optimal atomization performance by considering the dynamics of liquid jet.

## Introduction

The application of air blast atomizers covers a broad range of applications from propulsion systems and gas turbine combustors to agriculture and spray drying. The transverse injection of a spray into a high velocity air flow provides us with better droplet mixing and its design is better suited to slurry fuel mixture, as its simplistic design has less tendency to clog [1].

In abovementioned applications, the particular system performance may depend heavily on liquid atomization, spray penetration, particle size distribution and the mixing process. This paper is mainly concerned with particle size distribution of air blast atomizer. Liquid jet injection into a gaseous cross flow is highly complex and has been extensively studied numerically and experimentally. This significant amount of research can be variously classified. One of the reasonable classifications is based on the regimes which breakup mechanism may develop into cross flow, namely jet break up, jet penetration and spray structure [2]. Our focus is on jet breakup in order to approximate initial droplet diameter to initialize FSBM.

Flowing stream of air generates the shear force and pressure that deform the jet and induce entrainment. The jet is bent toward a direction parallel to the free stream and disintegration occurs by mechanism of surface striping and column fracture, namely primary breakup. The product of primary atomization then undergoes a secondary breakup.

Wu et al. and Inamura et al. investigated all three regimes for various jet exits and cross flow conditions in series of publication. Inamura et al. studied the structure of the disintegrating liquid jet in a cross flow. Their observation showed that major fracture developed on windward surface. They later modeled jet trajectories as well as droplet breakup successfully. Wu et al. investigated further into the breakup process of the liquid jet in subsonic cross flow. Their qualitative observation determined the liquid jet trajectory, column fracture, and surface breakup. Sallam et al. more recently studied the primary breakup of non-turbulent liquid jets by employing pulsed shadowgraph and pulsed holograph observations of primary breakup regime and also provided a review of the past literature on the atomization [3-7].

In literature, there are mainly two methods for characterizing particle distribution, i.e. empirical and analytical. Several empirical relationships have been proposed since 1930s, e.g., Rosin–Rammler, Weibull, Nukiyama–Tanasawa, log-normal, root-normal, and log-hyperbolic, etc.

Analytical approaches have been developed more recently and in general, are capable of calculating the correct shape of distribution. However, experimental input is still required [9]. There are three analytical methods available in order of time of development, i.e. Maximum Entropy (ME), discrete probability function (DPF) and Stochastic method.

Priori physical interpretation was relied on nondeterministic generation process, proposed by Sellens and Brzustowski. In this view, the distribution that maximizes the entropy function, within predefined constraints, is the most probable. Fine tuning the global constraints led to a reasonable modeling of drop size and velocity joint distribution [10, 11].

The second analytical approach is the discrete probability function (DPF) which combines nondeterministic portion of ME with deterministic part in order to alleviate the difficulties of ME by prior describing the primary breakup and then further modeling of probabilistic generation of droplets. Unlike ME method, it is limited to primary atomization. There are reviews on maximum entropy (ME) and discrete probability function (DPF) available in Babinsky and Sojka [8].

The more recent stochastic model is developed by Gorokhovski and Saveliev and Apte et al. Basically, it stems from cascade of uncorrelated breakage events described in Kolmogorov's scenario of the breakup of particles. The stochastic breakup model considers a cascade of droplet generation from parent drop via breakup that is independent of its parent drop size. Liu, Weixing and Zhou et al. employed the Kolmogorov's scenario to model initial droplet size down to the size of stable droplet and further studied fractal characteristic of the droplet size distribution [12-16].

Stochastic model has been successfully applied to primary and secondary atomization of high Weber number sprays, which are typical of air blast atomizers. Liu discussed the finite stochastic breakup model in detail as a variant of stochastic model. We aim to apply Finite stochastic break up model to droplet generation of secondary atomization regime of airblast plain liquid jet with circular array of six liquid jets and investigate its accuracy by comparing with experimental results. The droplet size distribution is expressed in terms of Rosin-Rammler distribution parameter N and Sauter mean diameter. The outline of this paper is as follows: The first section discusses FSBM description and input requirement followed by experimental setup configuration and finally the result and discussion section compares the experimental result and modelling approximation. The numerical result agreed well with experiment.

## 2. Numerical method

### 2.1 Finite stochastic model

Airblast atomizers with high Weber numbers lead to broad range of droplet sizes. For very large Weber numbers, droplet disintegration can be modelled by fractal nature of atomization. Zhou et al. has proposed a mathematical model of droplet splitting at uniformly distributed probability which is based on Fractal characteristics of the droplet size distribution. Liu et al. further develops this method and evolves it into a procedure for finite stochastic model. FSBM procedure entails the following assumption:

1. The net effect of breakup and aggregation process can be considered as the effect of reduced rate of breakup.
2. The probability of splitting a droplet into only two sub-droplet is defined with a breakup probability P(D). According to this probability, disintegration only occurs when droplet diameter is larger than maximum stable droplet size. Further complication of other operational or geometrical parameters can be employed into this probability in order to make it more than a simple trigger value.

3. If a droplet break up to two sub-droplets, then the mass ratio of a sub-droplet to the mother droplet is a random variable in the closed interval [a, 1 − a] with the uniform probability distribution, where a is the minimum mass ratio of a sub-droplet to the mother droplet, obviously $0<a<0.5$ [14].

In literature, the term generation is often introduced to denote the age of a droplet. As a mother drop split into sub droplets a new generation is created, therefore every gap during which a droplet may split defines a generation. Consecutive generations are analogues to cascade of droplet breakup events.

Stochastic methods have been successfully applied to primary and secondary atomization. In this paper, we are concerned with secondary atomization, therefore required parameters should be determined, i.e. Initial droplet diameter $D_0$, maximum stable droplet size Dc, minimum mass ratio of a sub-droplet to the mother droplet and droplet breakup probability P (D). The last two parameters are not readily related to physics of atomization and are changed in accordance to better agreement with experimental results.

### 2.1.1 Initial droplet size

Flowing stream of air generates the shear force and pressure that deform the column of jet and induce entrainment. The jet is bent toward a direction parallel to the free stream and disintegration occurs by mechanism of surface striping and column fracture, namely primary breakup [17]. Momentum-flux ratio *q*, is a determining factor in the penetration height of liquid jet and all existent relations are functions of this parameter. When the momentum of liquid jet is high, it sounds reasonable for the jet to withstand the cross stream momentum and penetrate further.

$$q = \frac{\rho U^2\big|_{Jet}}{\rho U^2\big|_{crossflow}} \qquad (1)$$

Becker and Hassa introduced a relation to measure trajectory of the penetration depth of the liquid jet for wide range of q 1<q<40 and pressure (1.5 bars <p<15 bars) [18].

$$\frac{z}{d} = 1.48 q^{0.42} \ell_n \left(1 + 3.56 \frac{x}{d}\right) \qquad (2)$$

When Reynolds of liquid jet, therefore turbulent intensity, is increased, interaction of the turbulent eddies within the liquid jet provides enough kinetic energy to cause surface breakup. The size of the drops formed via surface irregularities are analogous to the size of the drops formed via secondary atomization. The jet further penetrates into cross stream until liquid core fragments into ligaments and relatively big drops. In process of primary breakup, we are mainly concerned with column fracture which provides flow with particles that are most likely to further splitting in sequence of generation.

Primary breakup does not produce mono-size distribution. Nevertheless, we are mainly concerned with size distribution associated with column fracture. Sallam et al. proposed a correlation to measure SMD after turbulent primary breakup along the surface of turbulent round liquid jet:

$$\frac{SMD}{\Lambda} = 0.56 \left[ \frac{y}{\left(\Lambda We_{L\Lambda}^{0.5}\right)} \right]^{0.5} \qquad (3)$$

This relation provides us with a reasonable estimation of initial droplet diameter along turbulent liquid jet, particularly near the tip of penetration region.

### 2.1.2 The maximum stable droplet size

High *We* values indicate a high likelihood of liquid breakup. Lefebvre derived a relation for maximum stable droplet size by first equating the aerodynamic drag to the surface tension force. Maximum droplet size is [19]:

$$d_{\max} = \frac{8\sigma}{C_D \rho U_R^2} \tag{4}$$

Where $U_R$ is relative velocity. The relative velocity is nearly equal to cross flow air velocity. The inclined liquid jet is entirely confronted with cross flow air and the relative velocity in the cross stream direction is almost equal to cross flow velocity.

### 2.1.3 Fitting values

Numerical methods still require some experimental input to produce acceptable result and in this sense, it is more like elaborate empirical fit. Minimum mass ratio of a sub-droplet to the mother drop and droplet breakup probability are not readily related to physics of atomization. They should be defined with accordance to experimental result and be tuned to better agreement between experimental and numerical results.

Liu et al. educed that the minimum mass ratio of a sub-droplet to the mother droplet is a constant for several different air-blast atomizers and experimental conditions which also gives quite satisfactory results in current study. The breakup probability was also covered in same article. The following break up probability is proposed.

$$P(D) = \begin{Bmatrix} 0, & D \le D_c \\ 1, & D > D_c \end{Bmatrix} \tag{5}$$

This probability indicates implicitly that there exists no preferred length scale between mother drop and the maximum stable droplet. This assumption can be directly justified on the basis of fractal nature of atomization [13, 15, 20].

### 2.2 Simulation outline

Simulation procedure involves a series of steps. First, considerable number of particles in accordance with initial size distribution is provided, then Droplets are chosen based on the breakup probability which favors all droplets with diameter bigger than critical diameter, and finally chosen drops split into at most two daughter droplets with mass ratio of random value in the closed interval [a, 1 − a]. At this point, the particle system has undergone a generation.

This simulation is repeated for different values of “a” to obtain most consistent result with experiment. Rosin-Rammler distribution parameter is employed to represent particle distribution after each generation. Consecutive generations lead to fewer and fewer droplets having diameter bigger than maximum stable droplet. Therefore finally, no further splitting is possible for which exponent of Rosin-Rammler, N, is infinite and Sauter mean diameter is almost equal to maximum stable diameter.

### Experimental setup and condition

Experiment configuration is already used by Shafaee et al. Fig. 1 shows a schematic of the experimental setup comprising three main parts: the liquid (water) feed line, the compressed gas (air) line and the atomizer. The liquid feed line consists of five elements including a liquid reservoir with a capacity of 1.1 $m^3$ connected to the main tap water, a stainless steel mesh strainer hampering any possible tiny debris from the liquid flow, a liquid piston pump with a regulating pressure in the range of 0 to 50 bar capable of providing liquid flow rates up to 50 L/min, a needle valve for flow rate adjustment and a rotameter having a measurement range of 0 to 70 L/min with an accuracy of ±1%. The compressed air line is composed of three elements, respectively, including a pre-charged compressed air reservoir having a capacity of 50 L with a maximum allowable pressure of 140 bar and a mesh strainer followed by an air pressure regulator capable of reducing maximum pressure of 230 bar to a range of 0 to 15 bar. The air flow rates have been calculated based on an air anemometry procedure allowing the velocity and mean flow rate calculation for different air pressures. The compressed air pressure in addition to the air regulator is also monitored at a location in the vicinity of the atomizer inlet in order to ensure no air leakage in the air line. The injector is a two-fluid atomizer connected to the water and compressed air lines using an interface fixture. The fixture is designed to adapt the injector liquid and air entries to the corresponding lines on the setup in addition to mounting the injector in a holder for carrying out subsequent experimental tests. The atomizer and its picture are shown in Fig. 2. The compressed air flows through the middle part of the atomizer while the liquid feeds through an annular passage mounted on the atomizer periphery. The annular passage ends in six inclined holes which are equally arranged on the atomizer periphery with a sector angle of 60°. Each hole is also placed with an angle relative to the atomizer central axis without a swirl angle. The liquid lines simultaneously emerging from the holes are intercepted with the compressed air flow creating the primary atomization zone within the atomizer, which, in turn, results in the production of a spray of drops at the atomizer exit. The geometrical parameters of the investigated atomizers and crossflow Weber numbers are shown in Table 1. Utilizing Malvern Mastersizer X. Experimental measurement was carried out at the horizontal plane at a distance of 5cm from the orifice of atomizer [21, 22].

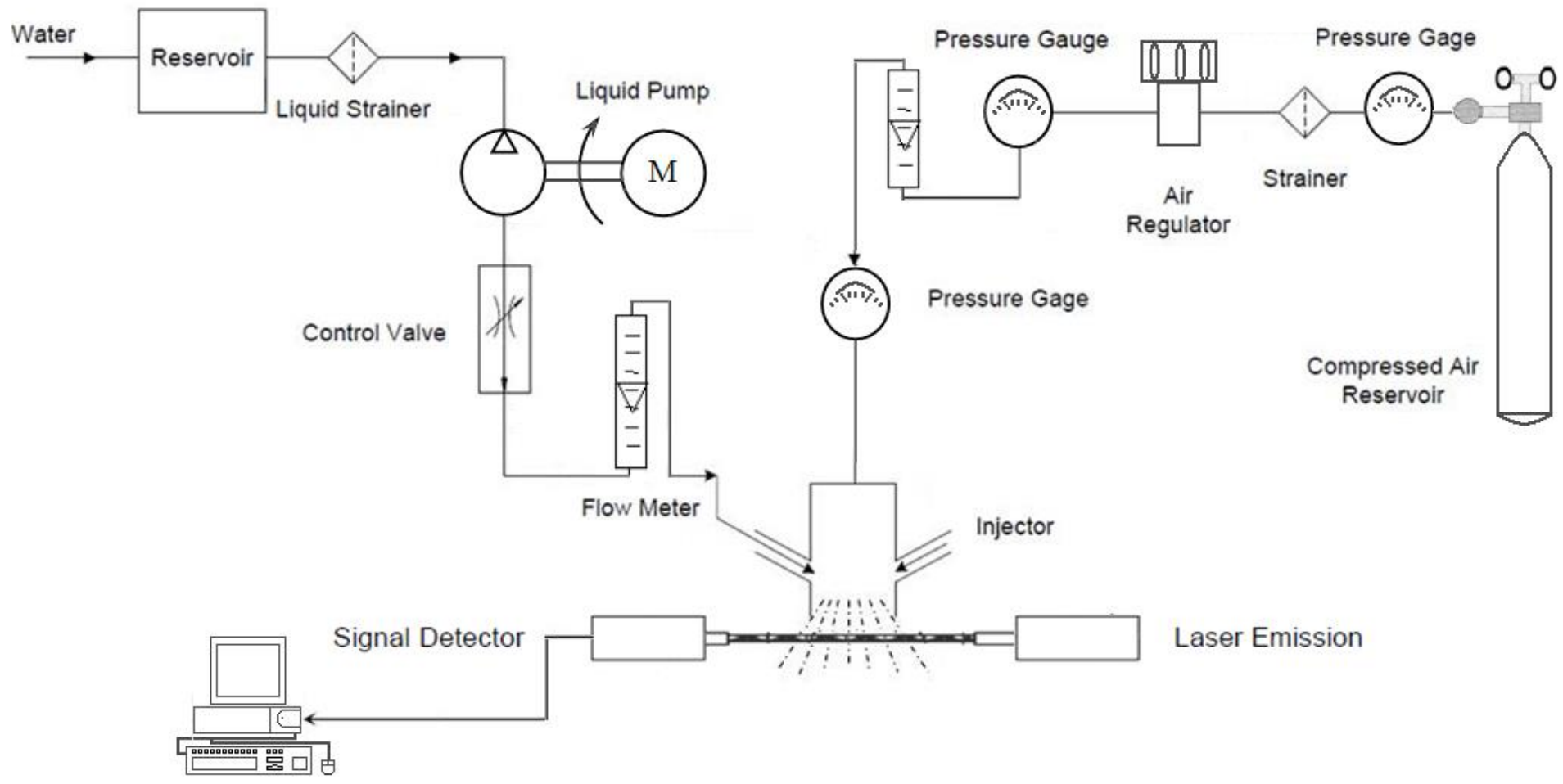

Fig. 1. Schematic view of the experimental setup.

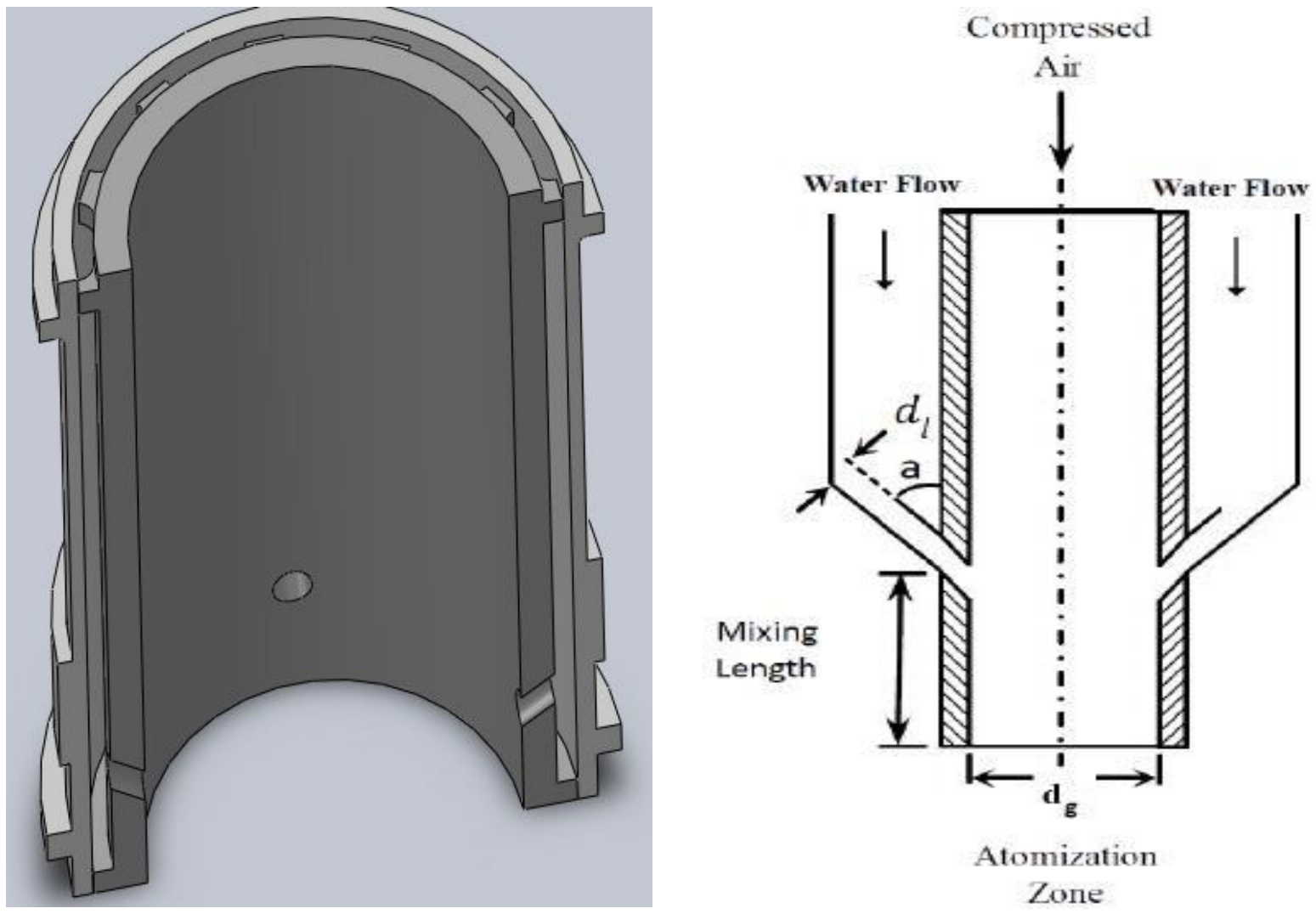

Fig. 2. The two-fluid atomizer used in this study, (a) 3-D view, (b) a schematic.

| No. | 1 | 2 | 3 | 4 | 5 | 6 | |
|---|---|---|---|---|---|---|---|
| $d_1$(mm) | 0.8 | 1 | 1.2 | 1.4 | 1.6 | 1.8 | |
| We | 727 | 1250 | 1750 | 2250 | 2500 | 2750 | 3000 |

Table 1. Experimental conditions of the investigated two-fluid atomizers.

## Results and discussion

Experimental results for test cases of table 1 is plotted in fig. 3 and 4. The variation of SMD with respect to Weber reveals a significant trend. Increasing Weber number which is achieved by increasing air stream velocity, reduces SMD. This trend is reasonable and not surprising. On the other hand, according to equation (4), we expect that maximum stable droplet diameter reduces as well. However, higher Weber numbers don't affect SMD very much, and in fact, SMD is insensitive to very high Weber numbers in favor of achieving more uniform particle distribution instead of keeping pace with smaller maximum stable diameters. SMD relates to Weber almost proportionally but end in plateau, as shown in fig. 3.

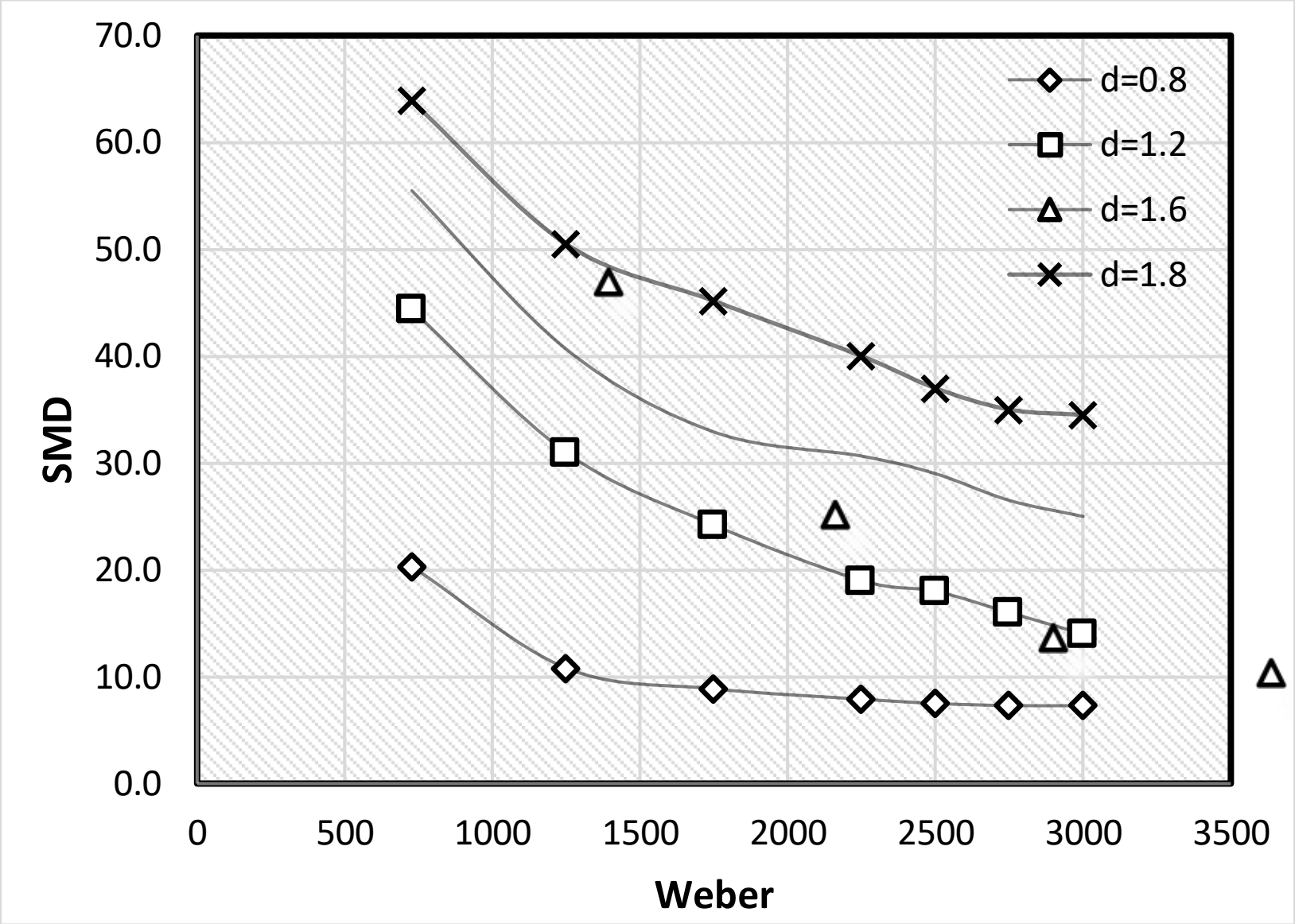

Fig. 3. The variation of SMD with respect to Weber number.

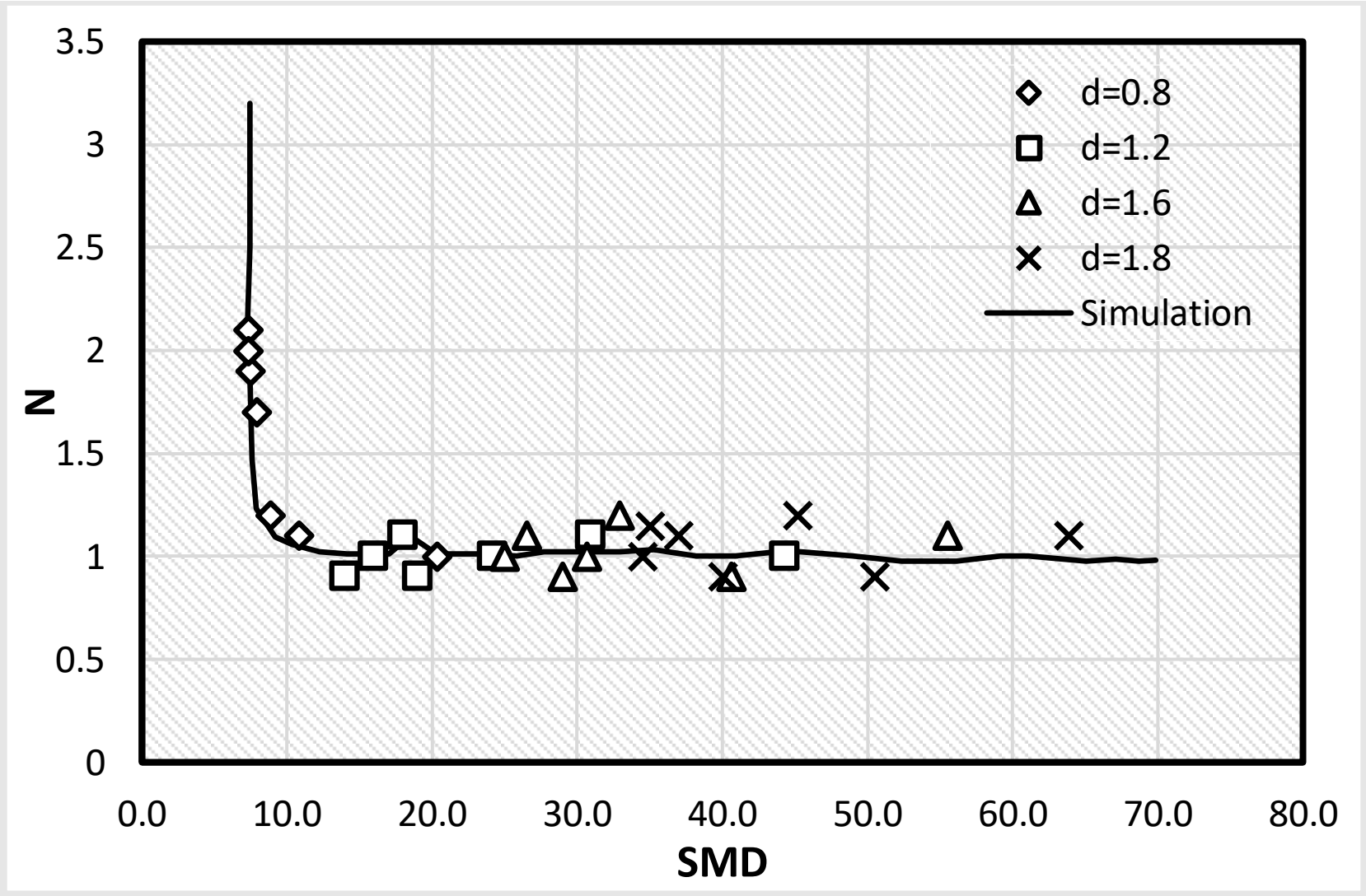

Fig. 4. The variation of N with respect to SMD.

As already mentioned, FSBM requires maximum stable droplet to be determined. Interestingly, maximum stable droplet associated with the beginning of the plateau in Fig. 4 is appropriate for describing all cases. In addition, minimum mass ratio of a sub-droplet to the mother droplet is the same as well, and equals to 0.14. Liu et al. also indicates that this value is constant for several different air-blast atomizers and experimental conditions. This is quite favorable property that reduces empirical input requirement, and therefore raises predictability value of method.

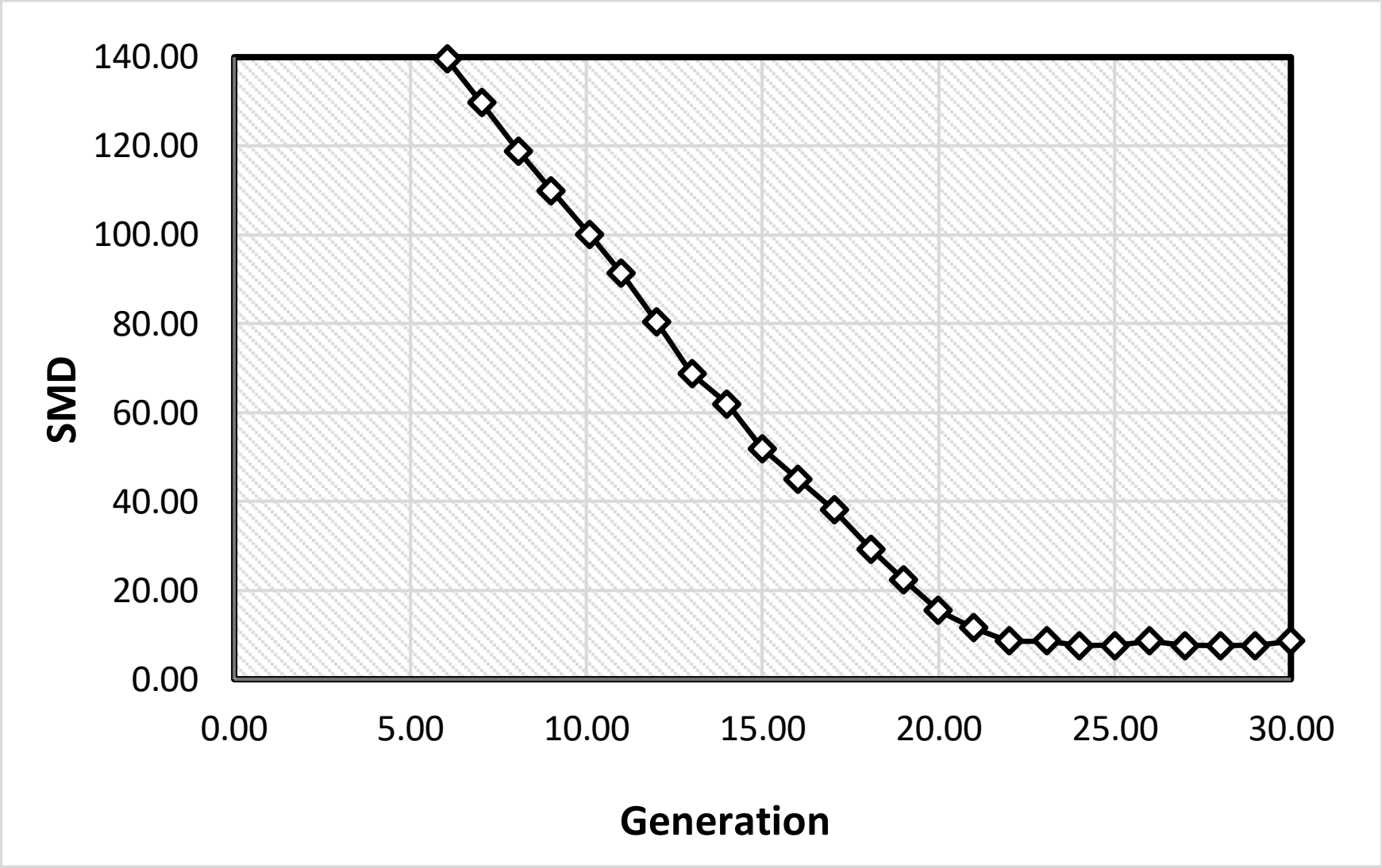

Fig. 5. Demonstration of simulation, the relation between SMD and generation

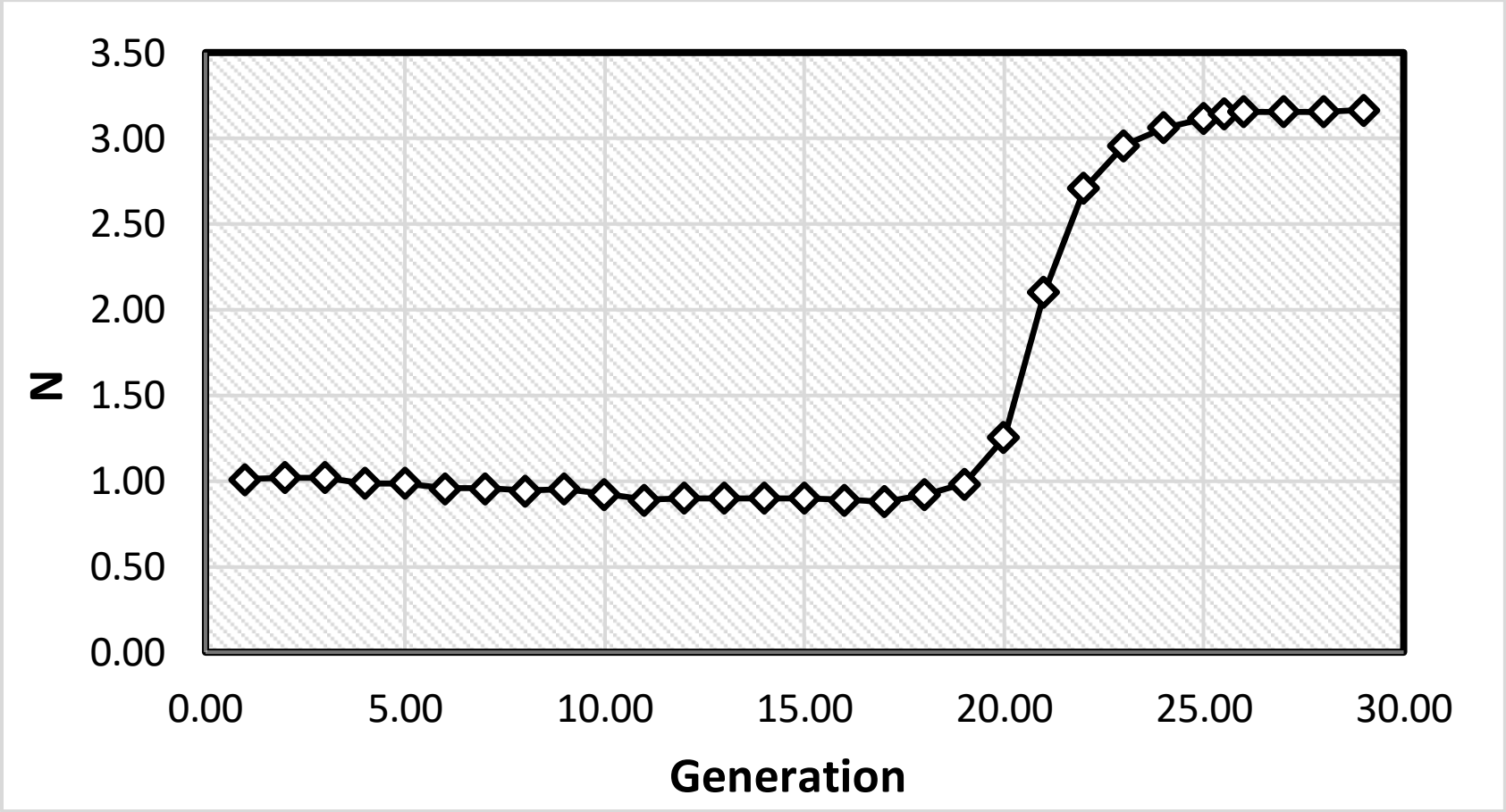

Fig. 6. Demonstration of simulation, the relation between N and generation

Figure 5 and 6 demonstrate FSBM results with the above mentioned values, in addition to $d_0$= 200 µm and $10^5$ particles. As generations are further created, decrement of SMD is reasonable, but abrupt change of N or plateau of SMD are informative and required to be explained. As generations are created, fewer and fewer particles are available as a source to feed into FSBM Algorithm. Liu et al. argued that as a result of this depletion, SMD asymptotically settles down, but on the other hand, N changes abruptly. N is calculated based on least square method which is known to be sensitive to outliers. The few remained particles with size bigger than Dc play the role of outliers for this method.

Figure 7 shows both experimental and numerical results with a good agreement. As Weber increases, SMD reduces and agrees well with simulation, and when there is no further reduction is achievable, particle distribution tends to become more uniform through higher values of N.

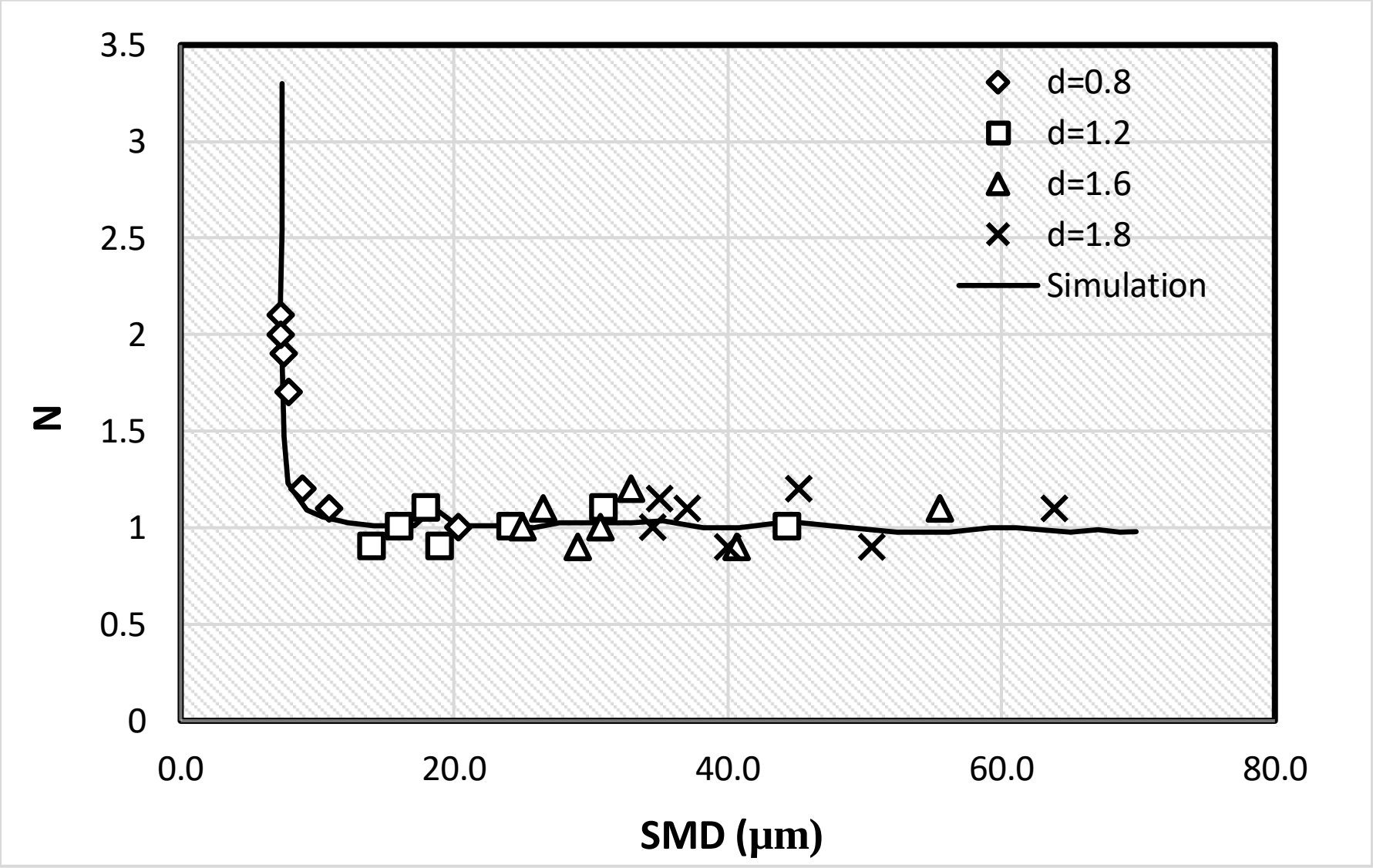

Fig. 7. Comparison between experimental and numerical results.

## Conclusion

Droplet size distribution in spray of plain-jet airblast atomizer with circular array of six inclined liquid jets is both experimentally and numerically studied. SMD and N of Rosin-Rammler are taken as characteristic parameter of distribution. The experiments are carried out for different Weber numbers and repeated for different jet diameters with constant Reynolds. The experiment shows that SMDs for spray of atomizers with different jet diameters basically follow same trend as Weber is increased and finally SMDs become insensitive to further increment of Weber. From this point forward, there is a significant change in N which is an index of uniformity of distribution.

FSBM capture this trend successfully, however, there are some empirical input required. The accuracy of FSBM only lies in the fact that there is no preferred length scale in transition of ligaments and drops to droplets. FSBM heavily relied on this fact and the input values are some kind of direction for start and end of the process.

Although empirical inputs are required, the basis for this algorithm does not show very sensitivity to change of atomizer parameters or operational conditions. This character makes FSBM very powerful and suitable for simulation for vast range of atomizers.

**NOMENCLATURE**

| | |
|---|---|
| $a$ | mass ratio of sub- droplet to the mother droplet |
| $D$ | droplet diameter ($\mu m$) |
| $l$ | mixing length(mm) |
| $d_l$ | liquid jet diameter (mm) |
| $D_{10}$ | average droplet diameter ($\mu m$) |
| $D_0$ | initial droplet diameter ($\mu m$) |
| $D_c$ | maximum stable droplet size ($\mu m$) |
| $I$ | modified Bessel function of first kind |
| $k$ | wave number |
| $K$ | modified Bessel function of second kind |
| $k_s$ | the wave number corresponding to the maximum growth rate |
| $N$ | measure of spread of droplet sizes |
| $Oh$ | Ohnesorge number, $Oh = \mu_f / \sqrt{\rho_f \sigma_f d_s}$ |
| $P(D)$ | breakup probability |
| $R$ | fraction of total volume |
| $u_r$ | relative velocity (m/s) |
| $We_g$ | Weber number, $We_g = \frac{\rho_g u_r^2 d_l}{\sigma}$ |
| $\alpha$ | $\frac{I_0}{I_1}$ |
| $\alpha_a$ | $\frac{K_0}{K_1}$ |
| $\omega$ | complex frequency |
| $\Gamma$ | Gamma function |
| $\mathrm{Re}_l$ | Reynolds number, $\mathrm{Re}_l = \frac{\rho_l u_r d_l}{\mu}$ |
| $\rho$ | density, kg/m$^3$ |
| $\mu$ | Viscosity, kg/(m.s) |
| $\sigma$ | surface tension, N/m |
| SMD | Sauter Mean Diameter $\mu$m |

**Subscripts**

| | |
|---|---|
| $g$ | gas |
| $l$ | Liquid |

## Reference

[1] Kihm, K.D., G.M. Lyn, and S.Y. Son, *ATOMIZATION OF CROSS-INJECTING SPRAYS INTO CONVECTIVE AIR STREAM.* 1995. 5(4&5): p. 417-433.

[2] Leong, M.Y., *Mixing of an airblast-atomized fuel spray injected into a crossflow of air*, V.G. McDonell and G.S. Samuelsen, Editors. 2000, Cleveland, Ohio: National Aeronautics and Space Administration Glenn Research Center.

[3] Inamura, T., et al., *Disintegration of liquid and slurry jets traversing subsonic airstreams.* Experimental Thermal and Fluid Science, 1993. 7(2): p. 166.

[4] Inamura, T. and N. Nagai, *Spray Characteristics of Liquid Jet Traversing Subsonic Airstreams.* Journal of Propulsion and Power, 1997. 13(2): p. 250-256.
[5] Inamura, T., *Trajectory of a Liquid Jet Traversing Subsonic Airstreams.* Journal of Propulsion and Power, 2000. 16(1): p. 155-157.
[6] Wu, P.-K., et al., *Breakup Processes of Liquid Jets in Subsonic Crossflows.* Journal of Propulsion and Power, 1997. 13(1): p. 64-73.
[7] Sallam, K.A., C. Aalburg, and G.M. Faeth, *Breakup of Round Nonturbulent Liquid Jets in Gaseous Crossflow.* AIAA Journal, 2004. 42(12): p. 2529-2540.
[8] Babinsky, E. and P.E. Sojka, *Modeling drop size distributions.* Progress in Energy and Combustion Science, 2002. 28(4): p. 303-329.
[9] Guildenbecher, D.R., C. López-Rivera, and P.E. Sojka, *Secondary atomization.* Experiments in Fluids, 2009. 46(3): p. 371-402.
[10] Shafaee, M., Abdehkakha, A., & Elkaie, A. (2014). Reverse analysis of a spiral injector to find geometrical parameters and flow conditions using a GA-based program. Aerospace Science and Technology, 39, 137-144.
[11] Shafaee, M., Sabour, M. H., Abdehkakha, A., & Elkaie, A. (2014). A visual study on the spray of gas-liquid atomizer. Indian J. Sci. Res, 1(2), 131-136.
[12] Gorokhovski, M.A. and V.L. Saveliev, *Analyses of Kolmogorov's model of breakup and its application into Lagrangian computation of liquid sprays under air-blast atomization.* Physics of Fluids, 2003. 15(1): p. 184-192.
[13] Apte, S.V., M. Gorokhovsk, and P. Moin, *LES of atomizing spray with stochastic modeling of secondary breakup.* International Journal of Multiphase Flow, 2003. 29: p. 1503-1522.
[14] Liu, H.-F., et al., *Prediction of droplet size distribution in sprays of prefilming air-blast atomizers.* Chemical Engineering Science, 2006. 61(6): p. 1741-1747.
[15] Zhou, W.X. and Z.H. Yu, *Multifractality of drop breakup in the air-blast nozzle atomization process.* Phys Rev E Stat Nonlin Soft Matter Phys, 2001. 63(1 Pt 2): p. 016302.
[16] Weixing, Z., et al., *Application of fractal geometry to atomization process.* Chemical Engineering Journal, 2000. 78(2–3): p. 193-197.
[17] Shafaee, M., Mahmoodzadeh, S., & Abdeh, A. (2016). Experimental and numerical investigation on airblast atomizer. Aerospace Knowledge and Technology Journal, 5(1), 41-53.

[18] Becker, J. and Hassa, C., Breakup and Atomization of a Kerosene Jet in Crossflow at Elevated
Pressure, Atomization and Sprays, 12(1–3), 2002, 49–67.
[19] Lefebvre, A.H., *Atomization and sprays*. 1989, New York; Washington ; London etc.: Hemisphere Pub. Corp. xi-421 p.
[20] Apte, S.V., et al., *Stochastic modeling of atomizing spray in a complex swirl injector using large eddy simulation.* Proceedings of the Combustion Institute, 2009. 32(2): p. 2257-2266.
[21] Shafaee, M., et al., *Effect of flow conditions on spray cone angle of a two-fluid atomizer*. Journal of Mechanical Science and Technology, 2011. 25(2): p. 365-369.
[22] Shafaee, M., et al., *An investigation on effect of geometrical parameters on spray cone angle and droplet size distribution of a two-fluid atomizer.* Journal of Mechanical Science and Technology, 2011. 25(12): p. 3047-3052.